\documentclass[aps,prb,twocolumn,showpacs]{revtex4-1}
\usepackage{bm}
\usepackage{graphicx}
\usepackage{amsmath,amsthm,amssymb}
\usepackage{dsfont}
\usepackage{soul}

\usepackage{color}

\begin{document}

\title{Spin resonance under topological driving fields}

\author{A.A. Reynoso,$^{1,2}$ J. P. Baltan\'as,$^3$ H. Saarikoski,$^4$ J. E. V\'azquez-Lozano,$^3$ J. Nitta,$^5$ and D. Frustaglia$^3$}
\affiliation{$^1$Instituto Balseiro and Centro At\'omico Bariloche, Comisi\'on Nacional de Energ\'ia At\'omica, 8400 Bariloche, Argentina}
\affiliation{$^2$Consejo Nacional de Investigaciones Cient\'ificas y T\'ecnicas (CONICET), Argentina}
\affiliation{$^3$Departamento de F\'isica Aplicada II, Universidad de Sevilla, E-41012 Sevilla, Spain}
\email{e-mail: frustaglia@us.es}
\affiliation{$^4$RIKEN Center for Emergent Matter Science (CEMS), Saitama 351-0198, Japan}
\affiliation{$^5$Department of Materials Science, Tohoku University, Sendai 980-8579, Japan}

\date{\today}

\begin {abstract}
We study the dynamics of a localized spin-1/2 driven by a time-periodic magnetic field that undergoes a topological transition. Despite the strongly non-adiabatic effects dominating the spin dynamics, we find that the field's topology appears clearly imprinted in the Floquet spin states through an effective Berry phase emerging in the quasienergy. This has remarkable consequences on the spin resonance condition suggesting a whole new class of experiments to spot topological transitions in the dynamics of spins and other two-level systems, from nuclear magnetic resonance to strongly-driven superconducting qubits.
\end{abstract}

\maketitle

\section{Introduction}

The manipulation of spin states by guiding fields in mesoscopic circuits offers several possibilities, from electron spin resonance\cite{SKGOKNSS13} to electron spin interferometry,\cite{NFSRN13} among others. This includes the prospects of spin control by geometric means, i.e., by making use of geometric phases\cite{BMKNZ03} that arise under the action of magnetic textures with a suitable topology.\cite{lyanda-geller} Recently, we have shown that guiding magnetic textures undergoing a topological transition can imprint this property in complex spin dynamics.\cite{SVBNNF15} More concretely, we reported electron transport simulations in spin interferometers subject to hybrid spin-orbit/magnetic textures showing a phase dislocation in the conductance as the distinct signature of a topological transition. This result is intriguing as the complexity of the spin dynamics near the critical point does not smooth the way for an intuitive picture of the transition in terms of geometric spin phases.

Here we address the problem of a localized spin subject to the action of time-periodic driving fields which are the time-dependent equivalent of the topological field textures studied in Ref. \onlinecite{SVBNNF15} for spin carriers.\cite{SVBNF16,G-LP12} This leads us to work within the Floquet framework, an area that has recently become very active due to the possibility of generating novel (topological) phases of matter.\cite{Kitagawa,Lindner,CiracFloquet,FrustagliaFloquet} We show that the topological characteristics of the guiding field have striking consequences on the spin resonance condition, leading to a definite inflection of the Bloch-Siegert shift\cite{BS40} at the critical point where the field undergoes a topological transition. We therefore propose corresponding resonance experiments as a proof of concept for topological transitions in the dynamics of spins-1/2 or any other two-level system as, e.g., strongly-driven superconducting qubits.
This approach has the advantage to overcome the difficulties present in spin-carrier implementations as those arising from disorder, multichannel transport and dephasing in hybrid spin-orbit/magnetic textures. Additionally, we develop a suitable theory that captures the topological signature left by the guiding field in terms of an emergent, effective Berry phase in the quasienergy of Floquet spin states. 

In Sec. \ref{sec-2} we introduce the concept of topological driving fields in Floquet theory. In Sec. \ref{sec-3} we present our main results concerning the topological imprints on spin resonances. In Sec. \ref{sec-4} we present a theory revealing complementary topological features in the Floquet spin states. Brief concluding remarks appear in Sec. \ref{sec-5}.  Technical details are discussed in Appendices \ref{AppA}, \ref{AppB} and \ref{AppC}.

\section{Spin dynamics under topological driving fields}
\label{sec-2}

\begin{figure}
\includegraphics[width=\columnwidth]{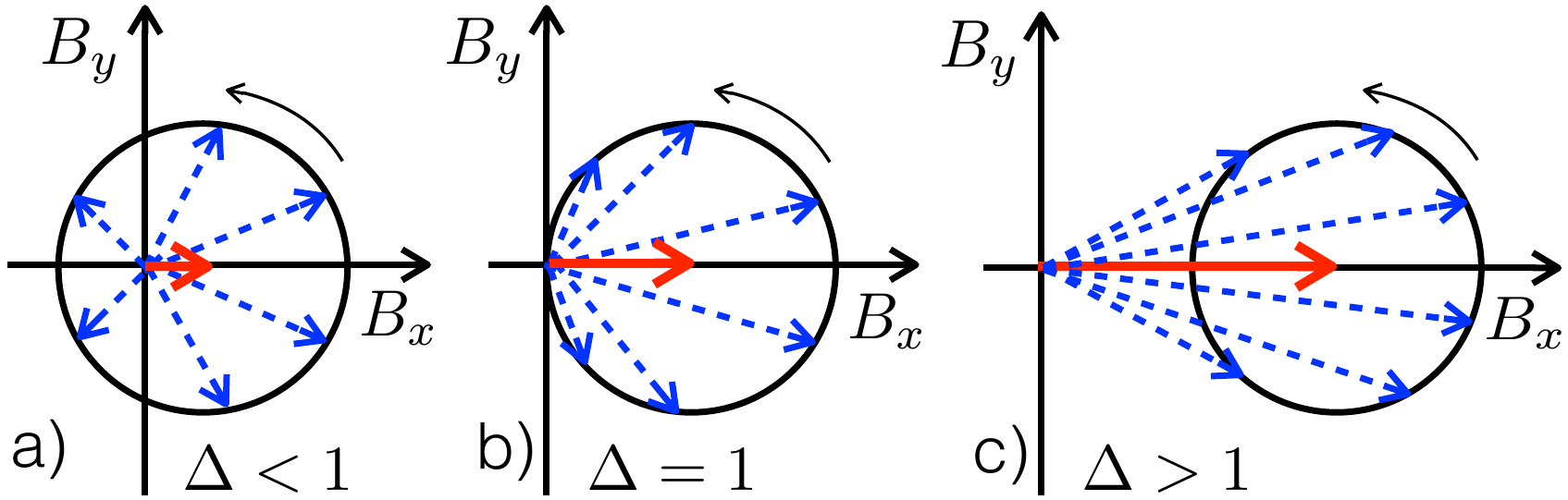}
\caption{Guiding field texture (dashed arrows) undergoing a topological transition at $\Delta=1$, from a rotating field ($\Delta < 1$) to an oscillating one ($\Delta > 1$). The field consist of a rotating driving ${\mathbf B}_1(t)$ plus a constant ${\mathbf B}_2$ (solid arrow), with $\Delta \equiv B_2/B_1$.
}
\label{fig-1}
\end{figure}

Consider a localized spin-1/2 guided by a time-dependent magnetic field consisting of two coplanar components: a rotating driving $\mathbf{B}_1(t)= B_1 (\cos \omega_0 t \ \hat{\mathbf x} + \sin \omega_0 t \ \hat{\mathbf y})$  and a constant $\mathbf{B}_2=B_2 \ \hat{\mathbf x}$ (see Fig. \ref{fig-1}). This is a Floquet problem described by the time-dependent Schr\"odinger equation
\begin{equation}
\mathcal{H}(t)|\Psi (t)\rangle=0
\label{floquet}
\end{equation}
with Floquet operator $\mathcal{H}(t) \equiv H(t)-i\hbar \ \partial/\partial t $ and Hamiltonian
\begin{equation}
H(t)= \frac{\hbar \omega_1}{2} \left( \cos \omega_0 t \ \sigma_x + \sin \omega_0 t \ \sigma_y \right) + \frac{\hbar \omega_2}{2} \sigma_x,
\label{H}
\end{equation}
where we have introduced the Larmor frequencies $\omega_{1,2}=\mu B_{1,2}/\hbar$, with $\mu$ the gyromagnetic ratio. The general solution of Eq. (\ref{floquet}) reads 
\begin{equation}
|\Psi(t)\rangle = \exp\left(-i \frac{\varepsilon t}{\hbar}\right) |\psi(t)\rangle,
\end{equation}
with quasienergy $\varepsilon$ and periodic Floquet spin state (FSS) $|\psi(t)\rangle$ satisfying the eigenvalue equation 
\begin{equation}
\mathcal{H}(t)|\psi(t)\rangle=\varepsilon |\psi(t)\rangle.
\label{floquet-2} 
\end{equation}
The FSSs have the form 
\begin{equation}
|\psi^+\rangle=\left(\begin{array}{c}\cos\frac{\theta}{2}e^{-i\varphi} \\ \sin\frac{\theta}{2} \end{array}\right),\quad |\psi^-\rangle=\left(\begin{array}{c}-\sin\frac{\theta}{2}e^{-i\varphi} \\ \cos\frac{\theta}{2} \end{array}\right),
\label{qstates}
\end{equation}
with time-dependent $\theta(t)$ and $\varphi(t)$. The spin dynamics resulting from Eq. (\ref{floquet-2}) can be rather complex depending on the parameter setting in the Hamiltonian (\ref{H}). As a rule, the instantaneous quantization axis of the FSSs (\ref{qstates}) does not point along the direction defined by the instantaneous field  $\mathbf{B}(t)=\mathbf{B}_1(t)+\mathbf{B}_2$. Instead, it typically presents a finite projection out of the field's plane characterized by a $\theta(t) \neq \pi/2$ for some $t$ in (\ref{qstates}). An exception to this rule is found in the limit of adiabatic spin dynamics where the instantaneous Larmor frequency of spin precession, $\omega (t) \equiv \sqrt{(\omega_2+\omega_1 \cos \omega_0 t)^2+(\omega_1 \sin \omega_0 t)^2}$, is much larger than the rotating field's frequency $\omega_0$.\cite{PFR03,berry} In this limit, the FSSs stay (anti)aligned with the instantaneous guiding field for all $t$. These adiabatic FSSs reduce to
\begin{equation}
|\psi^\uparrow \rangle=\frac{1}{\sqrt{2}}\left(\begin{array}{c} e^{-i\eta} \\ 1 \end{array}\right),\quad |\psi^\downarrow \rangle=\frac{1}{\sqrt{2}}\left(\begin{array}{c} - e^{-i\eta} \\ 1 \end{array}\right),
\label{adqstates}
\end{equation}
with $\eta(t)=\arctan\left[\sin \omega_0 t /(\cos \omega_0 t + \Delta)\right]$ and $\Delta = \omega_2/\omega_1$. 

The quasienergy of a FSS naturally splits into a dynamical and a geometrical contribution as 
\begin{equation}
\varepsilon^s= \frac{1}{T} \int_0^T \langle \psi^s| \mathcal{H}(t)|\psi^s \rangle \ {\rm d}t= \bar{E}^s-\frac{\hbar}{T}\phi^s_{\rm g},
\label{QE}
\end{equation}
where
\begin{eqnarray} 
\bar{E}^s &=& \frac{1}{T} \int_0^T \langle \psi^s| H(t)|\psi^s \rangle \ {\rm d}t \nonumber \\
&=& \frac{s}{T} \int_0^T \frac{\hbar \omega(t)}{2} \sin \theta \cos(\varphi-\eta) \ {\rm d}t
\label{ME}
\end{eqnarray}
is the mean energy and 
\begin{eqnarray} 
\phi^s_{\rm g} &=&  i \int_0^T \langle \psi^s| \frac{\partial}{\partial t}|\psi^s \rangle \ {\rm d}t \nonumber \\
&=& \frac{1}{2} \int_0^T (1+s\cos \theta) \frac{\partial \varphi}{\partial t} \ {\rm d}t = -\frac{1}{2} \Omega^s
\label{AAphase}
\end{eqnarray} 
is the geometric phase of the FSS with $s=\pm$, with $\Omega^s$ the corresponding solid angle ($\text{mod}[4\pi]$) subtended over the Bloch sphere during the spin evolution in one time period $T=2\pi/\omega_0$. For the adiabatic FSSs (\ref{adqstates}), this geometric phase reduces to a Berry phase.\cite{berry} Otherwise, it adopts the name of (non-adiabatic) Aharonov-Anandan (AA) phase.\cite{aharonov}

We notice in (\ref{H}) that the magnetic field undergoes a topological transition at $\Delta = 1$: a rotating field for $\Delta < 1$ (dominated by $\mathbf{B}_1(t)$ and enclosing the point of vanishing field in its round trip, Fig. \ref{fig-1}a) turns into an oscillating field for $\Delta > 1$ (dominated by $\mathbf{B}_2$ and leaving the point of vanishing field out of the loop, Fig. \ref{fig-1}c). At $\Delta = 1$, the total field $\mathbf{B}(t)$ vanishes whenever $\cos \omega_0 t = -1$, see Fig. \ref{fig-1}b. Such topological characteristics leave an imprint in the adiabatic FSSs (\ref{adqstates}): in a time period $T$ they acquire a geometric (Berry) phase $\phi_{\rm B} = \pi$ for $\Delta < 1$ while $\phi_{\rm B} = 0$ for $\Delta > 1$ [obtained after setting $\cos \theta =0$ in Eq. (\ref{AAphase})]. Topological transitions of this kind have been expected to show up in transport experiments with spin carriers relying on Berry-phase interference effects.\cite{lyanda-geller} However, this reasoning turns out to be oversimplified: in the vicinity of the transition point $\Delta = 1$, Fig. \ref{fig-1}b, the adiabatic condition can not be satisfied since the guiding field vanishes to reverse its direction and the spins are unable to follow it. Still, a more general approach has shown that field textures can leave a topological signature in electron spin transport simulations even far from the adiabatic regime.\cite{SVBNNF15} A numerical exploration of the time-dependent system modelled by Eq. (\ref{floquet}) reveals several topological imprints left by the guiding fields away from the adiabatic regime, as we discuss in Secs. \ref{sec-3} and \ref{sec-4}.

\section{Topological imprints on spin resonance}
\label{sec-3}

\begin{figure}
\includegraphics[width=\columnwidth]{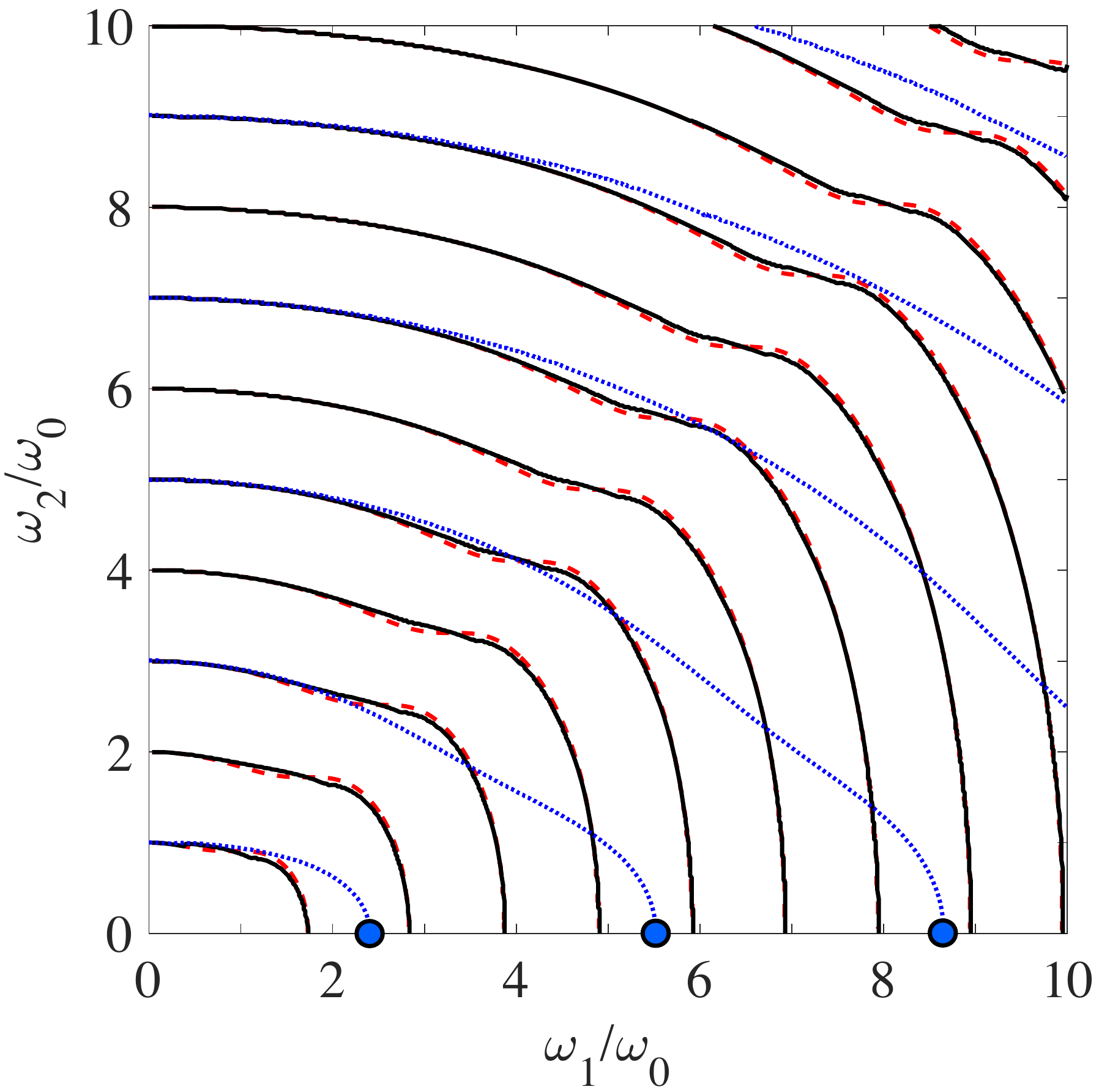}
\caption{
Solid lines: position of the resonances as a function of the guiding field's setting developing a Bloch-Siegert shift as the driving strength $\omega_1/\omega_0$ increases. The inflections along the diagonal $\omega_1=\omega_2$ (critical line $\Delta =1$) indicates a change in the guiding field's topology. Dashed lines: points of vanishing mean energy. Dotted lines: position of the resonances for a standard linear driving with trivial topology. The circles indicate the Stenholm's points.\cite{S72} 
}
\label{fig-2}
\end{figure}

A resonant transfer of energy between the driving field $\mathbf{B}_1(t)$ and the spins occurs whenever their expectation value along the uniform field $\mathbf{B}_2$ vanishes in a time-period average, i.e., whenever $\langle \langle \sigma_x \rangle \rangle_T=0$. Within the Floquet formalism, this resonant condition reads\cite{S65} 
\begin{equation}
\frac{\partial \varepsilon}{\partial \omega_2} =0, 
\label{resonances}
\end{equation}
which can be obtained by applying the Hellmann-Feynman theorem \cite{F39} to Eq. (\ref{floquet-2}) and gives the position of single- and multiple-photon processes. The solid lines in Fig. \ref{fig-2} show the position of the resonances according to Eq. (\ref{resonances}) [see Appendix \ref{AppA} for details on the numerical method]. For a weak driving $\omega_1/\omega_0 \ll 1$, resonances appear at integer values of $\omega_2/\omega_0$ due to the combined action of the oscillating components parallel and normal to the uniform field $\mathbf{B}_2$.\cite{S65} 
For larger driving amplitudes, the resonances shift their position to smaller (non integer) values of $\omega_2/\omega_0$ due to the Bloch-Siegert effect.\cite{BS40} Eventually, the resonance curves develop inflection points organized along the critical line $\Delta =1$. These inflections in the resonance profile are an actual signature of the topological transition undergone by the guiding field's texture at $\Delta =1$, imprinted in the resonant FSSs. The identification of such topological features in the Bloch-Siegert shift is the main result of this paper.

For a comparison, the dotted lines in Fig. \ref{fig-2} show the position of the resonances in the case of a standard linear driving owning a trivial topology $(\hbar \omega_1/2) \sin \omega_0 t \ \hat{\mathbf y}$ [to be compared with the topological driving of Eq. (\ref{H})]. These resonances present a usual Bloch-Siegert shift meeting the horizontal axis at the Stenholm's points (circles).\cite{S72} 

Moreover, the dashed lines in Fig. \ref{fig-2} show the points where the mean energy vanishes, running tight along the resonances (solid lines). This suggests that the condition of vanishing mean energy is a good approximation to the resonance condition (\ref{resonances}) for our topological driving. Consequently, in Sec. \ref{sec-4} we focus on the related topological features underlying at the level of the quasienergy, the mean energy and the geometric phase introduced in Eqs.~(\ref{QE})-(\ref{AAphase}). We notice that this approximation is not valid in general cases: it applies to our topological driving but not to linear drivings. In the latter case, the resonant condition (\ref{resonances}) coincides with the vanishing-mean-energy one only for weak drivings ($\omega_1/\omega_0 << 1$). 
 
\begin{figure*}
\includegraphics[width=18cm]{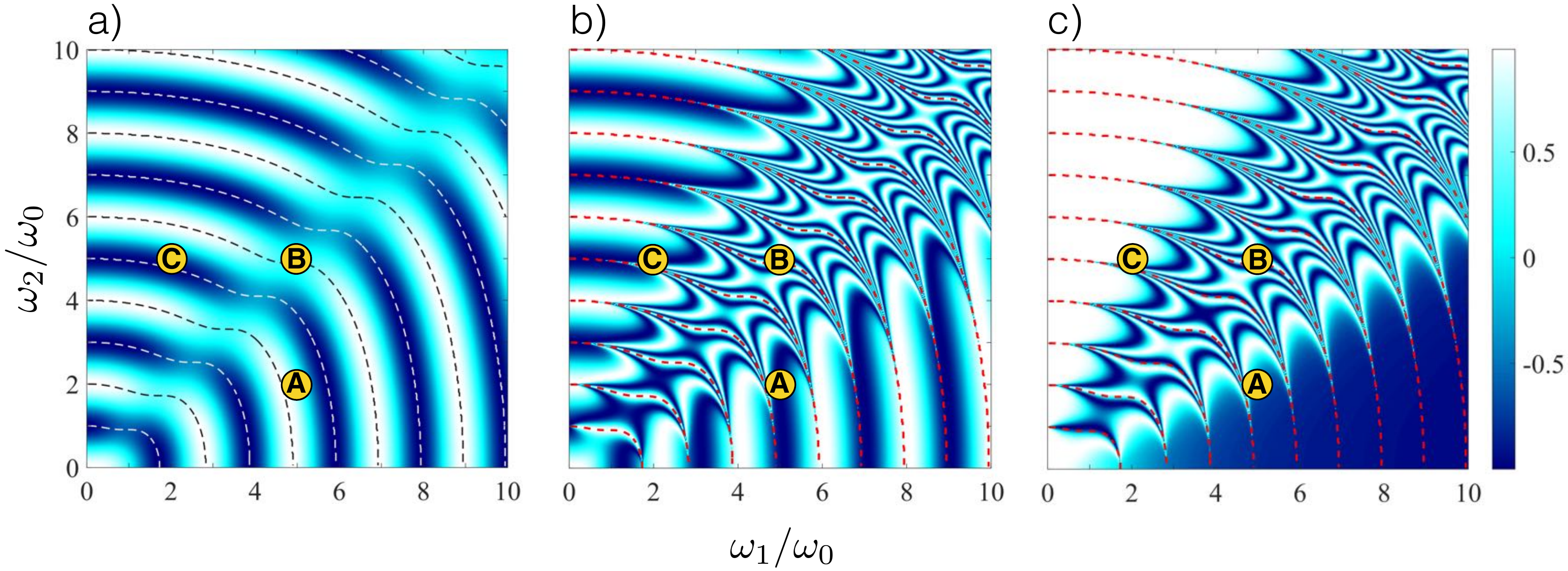}
\caption{
a) Response of the quasienergy $\varepsilon^+$ to the guiding field's setting in terms of $\cos(\varepsilon^+T/\hbar)$. The dislocation along the diagonal $\omega_1=\omega_2$ (critical line $\Delta =1$) indicates a change in the guiding field's topology. b) Mean energy $\bar{E}^+$ in terms of $\cos(\bar{E}^+T/\hbar)$. The complex pattern displayed near the diagonal is indicative of a non-adiabatic spin dynamics. c) Cosine of the AA geometric phase $\phi^+_{\rm g}$ displaying a complex pattern complementary to that  one shown by the mean energy in panel b). In all panels, the dashed lines indicate the points of vanishing mean energy $\bar{E}^+=0$. Spin textures of the Floquet state $| \psi^+\rangle$ corresponding to field settings A, B and C are shown in Fig. \ref{fig-4}. A study of the Floquet state $| \psi^- \rangle$ produces equivalent results.}
\label{fig-3}
\end{figure*}

\begin{figure}
\includegraphics[width=\columnwidth]{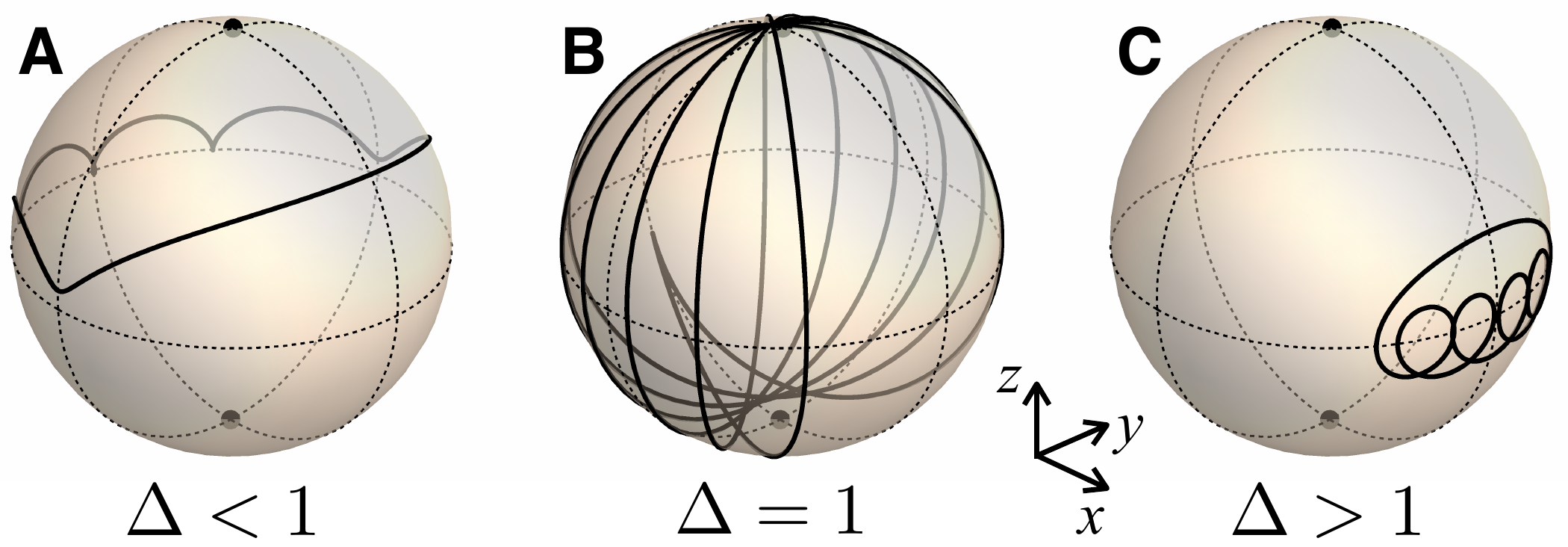}
\caption{Spin texture of the Floquet state $|\psi^+\rangle$  represented as a path over the surface of the Bloch sphere for different guiding field's settings (see points A, B and C in Figs. \ref{fig-3} and \ref{fig-5}).}
\label{fig-4}
\end{figure}

\section{Underlying topological features}
\label{sec-4}

In Fig. \ref{fig-3}a we depict the quasienergy (\ref{QE}) as a function of the guiding field's setting in Eq. (\ref{H}) [see Appendix \ref{AppA} for details on the numerical method]. More specifically, and without loss of generality, we plot $\cos (\varepsilon^+ T/\hbar)$ as a function of $\omega_1/\omega_0$ and $\omega_2/\omega_0$.\cite{note-1} There we observe a phase dislocation along the $\Delta =1$ diagonal as a sign of the quasienergy's response to the topology of the guiding field. This behaviour resembles the predictions for the conductance of semiconducting Rashba loops subject to similar field configurations (determined by the total phase acquired by spin carriers).\cite{SVBNNF15} It also recalls what expected for adiabatic spin dynamics according to Ref. \onlinecite{lyanda-geller}. However, as we show in the following, the spin dynamics is actually dominated by \emph{non-adiabatic} effects.   

In Figs. \ref{fig-3}b and \ref{fig-3}c we depict, respectively, the mean energy (in terms of the dynamical phase $\bar{E}^+ T/\hbar$) and the AA geometric phase by plotting $\cos (\bar{E}^+ T/\hbar)$ and $\cos \phi^+_{\rm g}$ as a function of the field's parameters [see Eqs.~(\ref{QE})-(\ref{AAphase}), and further details on numerics in Appendix \ref{AppA}]. There we notice the complex patterns displayed near the critical line $\Delta =1$, indicative of a strongly non-adiabatic spin dynamics. This is clearly illustrated by the FSS spin textures, Fig. \ref{fig-4}, adopting intricate shapes with rapidly changing solid angle in the critical region. That demonstrates the unsuitability of an adiabatic treatment for understanding the phase dislocation in the quasienergy (Fig. \ref{fig-3}a) as a response to the change in the guiding field's topology. 

\begin{figure}
\includegraphics[width=\columnwidth]{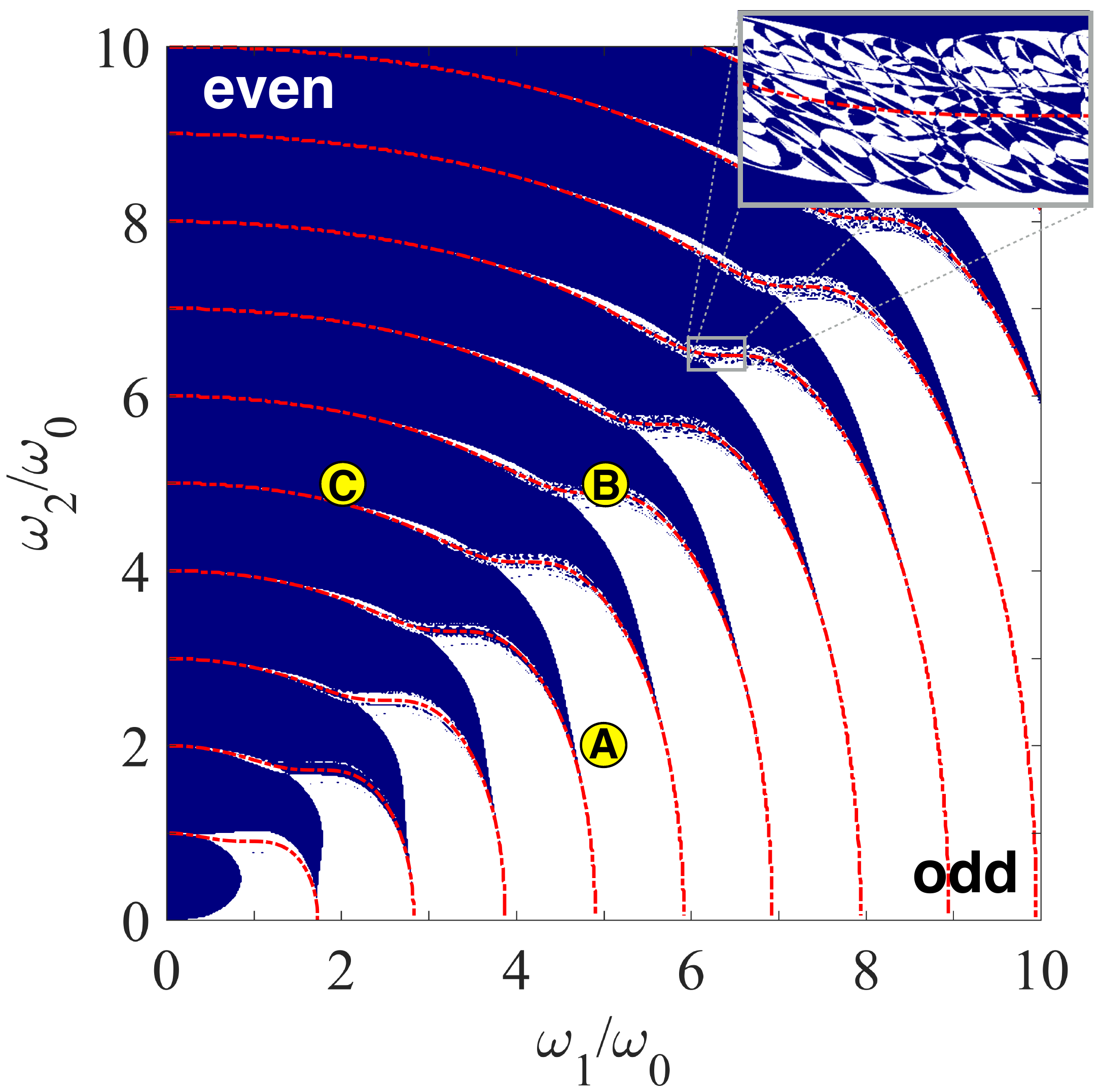}
\caption{Winding parity, showing a transition along the diagonal $\omega_1=\omega_2$ (critical line $\Delta =1$). Anomalous regions of fluctuating parity (inset) appear along the curves of vanishing mean energy (dashed), closely related to resonances. Spin textures of the Floquet states corresponding to points A, B and C are shown in Fig. \ref{fig-4}.}
\label{fig-5}
\end{figure}

To gain insight into the topological characteristics stamped on the spin dynamics, we introduce a modified version of a treatment first proposed in Ref. \onlinecite{ABRPR} for the study of (Berry) adiabatic phases in spin carriers, here adapted to the case of non-adiabatic spin dynamics (see Appendix \ref{AppB}). As a starting point, let us rewrite the Floquet operator as the sum of diagonal (d) and nondiagonal (nd) projections onto the non-adiabatic FSS basis (\ref{qstates}), i.e., $\mathcal{H}(t) = \mathcal{H}_{\rm d}(t) + \mathcal{H}_{\rm nd}(t)$. By the sole definition of Floquet state, it must hold $\mathcal{H}_{\rm nd}(t) \equiv 0$ [and, therefore, $\mathcal{H}(t) \equiv \mathcal{H}_{\rm d}(t)$]. As a consequence, the FSSs are constrained to satisfy 
\begin{equation}
\omega(t) \cos\theta \cos(\varphi-\eta)+\sin\theta \frac{\partial \varphi}{\partial t}=0, 
\label{real-ndfloquet-0}
\end{equation}
as shown in the Appendix \ref{AppB}. From Eqs. (\ref{ME})-(\ref{real-ndfloquet-0}) we can rewrite the quasienergy (\ref{QE}) as (see Appendix \ref{AppC})
\begin{equation}
\varepsilon^s = -\frac{s\hbar}{2T}\int_0^T \frac{1}{\cos \theta} \frac{\partial \varphi}{\partial t} \ {\rm d}t-  \frac{\hbar}{T} \ell\pi
\label{QE-2}
\end{equation}
thanks to the cancelation of the term proportional to $\cos \theta$ in Eq. (\ref{AAphase}). We recognize two contributions to the quasienergy in Eq. (\ref{QE-2}): a smooth dynamical term proportional to $1/\cos \theta$ [from Eq. (\ref{real-ndfloquet-0}) we see that this term does not diverge for vanishing $\cos \theta$] plus a topological one determined by the integer number 
\begin{equation}
\ell = \frac{1}{2\pi}\int_0^T \frac{\partial \varphi}{\partial t} \ {\rm d}t, 
\label{wn}
\end{equation}
accounting for the windings gathered by the FSSs around the north pole of the Bloch sphere. 
This is illustrated by the spin textures of Fig. \ref{fig-4}A with $\ell=1$ ($\Delta < 1$) and Fig. \ref{fig-4}C with $\ell=0$ ($\Delta > 1$).
We further notice from Eq. (\ref{QE-2}) that 
\begin{equation}
\cos(\varepsilon^s T/\hbar)= (-1)^\ell \cos(\phi_{\rm d}^0),
\end{equation}
with $\phi_{\rm d}^0 = \int_0^T \frac{1}{2\cos \theta} \frac{\partial \varphi}{\partial t} \ {\rm d}t$ the smooth contribution to the dynamical phase. Thus, a parity transition in the winding number $\ell$ at $\Delta = 1$ would explain the phase dislocation shown by the quasienergy in Fig. \ref{fig-3}a (together with the results reported in Ref. \ \onlinecite{SVBNNF15}) in terms of an effective Berry phase $\phi_{\rm B}^{\rm eff} \equiv \ell \pi$ recalling the geometric phase acquired by the adiabatic FSSs (\ref{adqstates}). Still, the actual existence of complex spin textures with singular derivative $\partial \varphi / \partial t$ (corresponding to a spin passing over the poles of the Bloch sphere as in Fig. \ref{fig-4}B) complicates the analysis. In Fig. \ref{fig-5} we represent the winding parity (i.e., the parity of the winding number $\ell$) as a function of the guiding field's configuration. As expected, we find opposite dominating parities on different sides of the critical line: even (dark) for $\Delta > 1$ and odd (white) for $\Delta < 1$. Interestingly, anomalous regions of fluctuating parity appear along the lines corresponding to a vanishing mean energy (dashed lines in Fig. \ref{fig-5}). Similar winding-number characteristics have been identified recently in spin carriers subject to magnetic textures.\cite{SVBNNF15,YGOC16} The ultimate reason for that is the proximity of spin resonances (as shown in Fig.  \ref{fig-2}) developing complex spin textures. Such anomalies should not be necessarily understood as noisy phases in the parameter space since they actually show a definite mosaic structure in finer scales (inset in Fig. \ref{fig-5}). The presence of anomalies ultimately indicates that the winding number $\ell$, while otherwise useful, is not an optimal indicator of the topological transition. The quest for an indicator that fully captures the topological origin of the phase dislocation in the quasienergy at $\Delta =1$, Fig. \ref{fig-3}a, remains open.

\section{Concluding remarks}
\label{sec-5}

We have studied the topological imprints left by an external driving in the dynamics of a localized spin-1/2. We find that the topological transition undergone by a guiding magnetic field emerges in the resonance spectrum of the spin system as a distinct inflection in the Bloch-Siegert shift. This remarkable feature opens a door to a new family of resonance experiments in two-level systems as an alternative to previous attempts to observe real-space topological transitions based on spin-carrier interferometry in mesoscopic loops,\cite{SVBNNF15} the realization of which presents technical difficulties up to date. The possibilities run from nuclear magnetic resonance (NMR) to strongly-driven superconducting qubits (SCQs). In NMR, these effects can be demonstrated experimentally by shaping radio frequency pulses for generating the suitable driving Hamiltonian in the rotating frame of the nuclear spins.\cite{WS88} The predicted effects might be significant for the design of shaped pulses for robust control of quantum systems.\cite{SA16} As for SCQs,\cite{OYLBLO05} we notice that high-order multiphoton interferometry has been demonstrated successfully and the systems can readily be adapted to the topological driving fields proposed here.\cite{note-2} All results extend to any other two-level system liable to equivalent drivings (as, e.g. adapted versions of Landau-Zener-St\"uckelberg interferometric setups \cite{LZS}).

For the topological driving considered here, we found that the resonance condition can be approximated by the condition of vanishing mean energy. This has permitted us to shift our analysis towards the dynamical and geometrical contributions to the quasienergy. As a consequence, we ended up with a description of the topological transition imprinted in the spin dynamics in terms of an effective Berry phase. This description, even when inexact, captures the essential features of the problem.

\begin{acknowledgments}
This work was supported by Project No. FIS2014-53385-P (MINECO, Spain) with FEDER funds and by Grants-in-Aid for Scientific Research (C) No. 26390014, for Specially Promoted Research No. H1505699, and for Scientific Research on Innovative Areas No. JP15K21717 (Japan Society for the Promotion of Science). AAR thanks the hospitality of the Departamento de F\'isica Aplicada II, Universidad de Sevilla. We thank G. \'Alvarez and W. D. Oliver for useful comments on NMR and SCQs, respectively.
\end{acknowledgments}

\appendix

\section{Numerical method.}
\label{AppA}

We solve Eq. (\ref{floquet-2}) by the customary procedure of expanding in Fourier series the Hamiltonian (\ref{H}), $H(t)=\sum_n H^{(n)} e^{i n \omega_0 t}$, and the FSS, $|\psi(t)\rangle=\sum_n |\psi^{(n)}\rangle e^{i n \omega_0 t}$. This leads to the infinite set of equations: $\sum_m H^{(n-m)}|\psi^{(m)}\rangle= (\varepsilon-n\hbar\omega_0) |\psi^{(n)}\rangle$ with $n$ integer. We solve the eigenvalue problem for $\varepsilon$ and $\{|\psi^{(n)}\rangle\}$ by restricting to a finite set of equations with $|n|\leq n_{max}$. We chose $n_{max}=120$ and checked that for the strongest simulated driving amplitudes the results are unaffected by the truncation. 

The mean energies are obtained from the Fourier decomposition of the Floquet states as $\bar{E}=\varepsilon-\hbar\omega_0\sum_n n |\langle\psi^{(n)} |\psi^{(n)}\rangle|^2$, where $\phi_{\rm g}=2\pi \sum_n n |\langle\psi^{(n)} |\psi^{(n)}\rangle|^2$ are the corresponding AA geometric phases.

\section{Non-adiabatic spin dynamics under periodic driving.}
\label{AppB}

We introduce a modified version of a treatment first proposed in Ref. \onlinecite{ABRPR} for the study of adiabatic (Berry) phases in spin carriers, here adapted to the case of non-adiabatic spin dynamics in periodic, time-dependent fields.\cite{BSRF17} Let us rewrite the Floquet operator of Eq. (\ref{floquet}) as the sum of diagonal (d) and nondiagonal (nd) projections onto the non-adiabatic FSS basis (\ref{qstates}), i.e., $\mathcal{H}(t) = \mathcal{H}_{\rm d}(t) + \mathcal{H}_{\rm nd}(t)$. To this aim, we define instantaneous projectors on the corresponding subspaces given by
\begin{equation}
{\mathcal P}_{\pm}(t)=\frac{1\pm\hat{\mathbf{l}}(t)\cdot\boldsymbol{\sigma}}{2},
\end{equation}
with $\hat{\mathbf l}(t)=\sin \theta(t) \cos \varphi(t) \hat{\mathbf x}+\sin \theta(t) \sin \varphi(t) \hat{\mathbf y}+\cos \theta(t) \hat{\mathbf z}$ the unit vector defining the instantaneous quantization axis of the FSSs at time $t$. We stress that $\hat {\mathbf l}(t)$ generally differs from the guiding field's axis $\hat {\mathbf n}(t)=\cos \eta(t) \hat{\mathbf x}+ \sin \eta(t) \hat{\mathbf y}$ in the non-adiabatic regime. We further notice that, by definition of Floquet states, it holds
\begin{eqnarray}
\mathcal{H}_{\mathrm d}(t)&=&{\mathcal P}_{+}{\mathcal H}{\mathcal
  P}_{+}+{\mathcal P}_{-}{\mathcal H}{\mathcal P}_{-}\equiv\mathcal{H}(t), \label{dfloquet}\\
\mathcal{H}_{\mathrm{nd}}(t)&=&{\mathcal H}-{\mathcal H}_{\mathrm d}={\mathcal
  P}_{+}{\mathcal H}{\mathcal P}_{-}+{\mathcal P}_{-}{\mathcal
  H}{\mathcal P}_{+} \equiv 0.
  \label{ndfloquetzero}
\end{eqnarray}
The constraint imposed by Eq. \ (\ref{ndfloquetzero}) establishes a definite link between the magnetic texture and the non-adiabatic FSS texture that eventually leads to the identification of an effective Berry phase imprinted by the guiding field. We can make the most of the formal identities (\ref{dfloquet}) and (\ref{ndfloquetzero}) by noticing that the projectors ${\mathcal P}_{\pm}$ do not commute with $i \hbar \partial/\partial t$ (which therefore mixes the FSS subspaces). By following Refs.~\onlinecite{ABRPR} and \onlinecite{FR01}, we introduce an operator ${\mathcal A}(t)$ responsible for the transitions between the subspaces associated with ${\mathcal P}_{+}$ and ${\mathcal P}_{-}$ while $i \hbar \partial/\partial t-{\mathcal A}$ acts only within each subspace. This is accomplished without ambiguity by defining
\begin{equation}
{\mathcal A}=i \hbar \frac{\partial}{\partial t}-{\mathcal P}_{+}\left(i \hbar \frac{\partial}{\partial t}\right){\mathcal P}_{+}-{\mathcal P}_{-}\left(i \hbar \frac{\partial}{\partial t}\right){\mathcal P}_{-},
\end{equation}
which verifies $[i \hbar \partial/\partial t-{\mathcal A},{\mathcal P}_{\pm}]=0$ and
${\mathcal P}_{\pm}{\mathcal A}{\mathcal P}_{\pm}=0$. By taking into account the
projectors' properties ($\mathcal{P}^2_{\pm}=\mathcal{P}_{\pm}$ and
$\mathcal{P}_{+}+\mathcal{P}_{-}=\mathds{1}$), one obtains 
\begin{eqnarray}
\mathcal{H}_{\mathrm{d}}(t)&=& \mu(\mathbf{B}\cdot\hat{\mathbf{l}})(\hat{\mathbf{l}}\cdot\boldsymbol{\sigma})- \left(i \hbar \frac{\partial}{\partial t}-{\mathcal A}\right),
\label{dfloquet-2} \\
\mathcal{H}_{\mathrm{nd}}(t)&=&\mu \left[\mathbf{B}\cdot\boldsymbol{\sigma}-(\mathbf{B}\cdot\hat{\mathbf{l}})(\hat{\mathbf{l}}\cdot\boldsymbol{\sigma})\right ] -{\mathcal A} \equiv 0,
\label{ndfloquet-2}
\end{eqnarray}
where we have dropped the dependence on $t$ when convenient for ease in notation. Moreover, the explicit evaluation of $\mathcal{A}$ produces
\begin{equation}
\mathcal{A}=-\frac{i\hbar}{2}(\hat{\mathbf{l}}\cdot\boldsymbol{\sigma})\frac{\partial}{\partial t}(\hat{\mathbf{l}}\cdot\boldsymbol{\sigma}).
\end{equation} 
We now rewrite $\mathcal{H}_{\mathrm{d}}$ and $\mathcal{H}_{\mathrm{nd}}$ in the FSS basis by introducing the instantaneous unitary operator 
\begin{equation}
\mathcal{U}(t)=\left(\begin{array}{cc}\cos\frac{\theta}{2}e^{i\varphi} & \sin\frac{\theta}{2}  \\ -\sin\frac{\theta}{2}e^{i\varphi} & \cos\frac{\theta}{2} \end{array}\right)
\end{equation}
such that
\begin{equation}
\mathcal{U}(t)|\psi^+\rangle=\left(\begin{array}{c} 1 \\ 0 \end{array}\right),\quad\mathcal{U}(t)|\psi^-\rangle=\left(\begin{array}{c} 0 \\ 1 \end{array}\right).
\end{equation}
We first notice that 
\begin{equation}
\mathcal{U}\mathcal{A}\mathcal{U}^{\dagger}=\left(\begin{array}{cc} 0 & a_{\rm g}^{+-}  \\ a_{\rm g}^{-+} & 0 \end{array}\right),
\end{equation}
where  $a^{s\bar{s}}_{\rm g} (t)= -\hbar(\sin \theta  \ \partial \varphi/\partial t +i \ s \  \partial \theta/\partial t)/2$ plays the role of a geometric mixing. Moreover, 
\begin{equation}
\mathcal{U} \left (i \hbar \frac{\partial}{\partial t}-\mathcal{A} \right)\mathcal{U}^{\dagger}=\left(\begin{array}{cc} i \hbar \frac{\partial}{\partial t}+A_{\mathrm{g}}^{+} & 0  \\ 0 & i \hbar \frac{\partial}{\partial t}+A_{\mathrm{g}}^{-} \end{array}\right),
\end{equation}
where
\begin{equation}
A_{\mathrm{g}}^{s}(t)=\frac{\hbar}{2}(1+s\cos\theta) \frac {\partial \varphi} {\partial t}
\end{equation}
is responsible for the emergence of non-adiabatic geometric phases (see Appendix \ref{AppC}). Back to the Floquet operator, we find
\begin{equation}
\mathcal{U}\mathcal{H}_{\mathrm{d}}\mathcal{U}^{\dagger}=\left(\begin{array}{cc}\mathcal{H}^{+} & 0  \\ 0 & \mathcal{H}^{-} \end{array}\right),
\end{equation}
with
\begin{equation}
\mathcal{H}^{s}(t)=s \ \mu \mathbf{B}(t)\cdot\hat{\mathbf{l}}(t)- \left(i \hbar \frac{\partial}{\partial t}+A_{\mathrm{g}}^s(t)\right).
\label{dfloquet-elements}
\end{equation}
The first term in Eq. (\ref{dfloquet-elements}) is the effective Zeeman splitting resulting from the instantaneous projection of the non-adiabatic FSSs along the guiding field, which can be also expanded as $\hbar \omega(t) \left(|\langle \psi^s|\psi^\uparrow \rangle|^2-|\langle \psi^s|\psi^\downarrow \rangle|^2\right)/2$ with $s=\pm$. Similarly, we find 
\begin{equation}
\mathcal{U}\mathcal{H}_{\mathrm{nd}}\mathcal{U}^{\dagger}=\left(\begin{array}{cc} 0 & \mathcal{H}^{\pm}  \\ \mathcal{H}^{\mp} & 0 \end{array}\right),
\label{ndfloquet-elements}
\end{equation}
with
\begin{eqnarray}
\mathcal{H}^{s\bar{s}}&=& \frac{\hbar \omega(t)}{2} \ \left(\langle \psi^s |\psi^\uparrow \rangle \langle \psi^\uparrow |\psi^{\bar{s}} \rangle - \langle \psi^s|\psi^\downarrow \rangle \langle \psi^\downarrow |\psi^{\bar{s}} \rangle \right) \nonumber \\
&-& a^{s\bar{s}}_{\rm g}(t)\equiv 0. \label{ndfloquet-elements}
\end{eqnarray}
The evaluation of the real and the imaginary parts of Eq. \ (\ref{ndfloquet-elements}) leads to the following identities, respectively:
\begin{eqnarray}
\omega(t) \cos\theta \cos(\varphi-\eta)+\sin\theta \frac{\partial \varphi}{\partial t}&=&0, \label{real-ndfloquet} \\
\omega(t) \sin(\varphi-\eta)+\frac{\partial \theta}{\partial t}&=&0. 
\label{imag-ndfloquet}
\end{eqnarray}
These equations reveal the geometric constraints satisfied by the FSSs under the action of a driving represented by $\omega(t)$ and $\eta(t)$. 

\section{Topological phases and quasienergy.}
\label{AppC}

From Eq. (\ref{dfloquet-elements}), we can now rewrite Eqs. (\ref{QE})-(\ref{AAphase}) as $\varepsilon^s= \langle \mathcal{H}^s(t)\rangle = \bar{E}^s-\hbar \phi^s_{\rm g}/T$ with
\begin{eqnarray}
\bar{E}^s &=& \frac{s}{T}\int_0^T \ \mu \mathbf{B}(t)\cdot\hat{\mathbf{l}}(t) \ {\rm d}t \nonumber \\ 
&=& \frac{s}{T} \int_0^T \frac{\hbar \omega(t)}{2} \sin \theta \cos(\varphi-\eta) \ {\rm d}t,\label{ME-2} \\ 
\phi^s_{\rm g} &=& \frac{1}{\hbar}\int_0^T \ A_{\mathrm{g}}^s(t) \ {\rm d}t \nonumber \\
&=& \frac{1}{2} \int_0^T (1+s\cos \theta) \frac{\partial \varphi}{\partial t} \ {\rm d}t. 
\label{AAphase-2}
\end{eqnarray}
From Eq. (\ref{real-ndfloquet}), we notice that the mean energy (\ref{ME-2}) takes the form
\begin{equation}
\bar{E}^s = \frac{s\hbar}{2T}\int_0^T \left(\cos \theta - \frac{1}{\cos \theta}\right) \frac{\partial \varphi}{\partial t} \ {\rm d}t.
\label{ME-3}
\end{equation}
The quasienergy reduces then to
\begin{equation}
\varepsilon^s = -\frac{s\hbar}{2T}\int_0^T \frac{1}{\cos \theta} \frac{\partial \varphi}{\partial t} \ {\rm d}t-  \frac{\hbar}{T} \ell \pi,
\label{QE-3}
\end{equation}
thanks to the cancellation of the terms proportional to $\cos \theta$ in Eqs. (\ref{AAphase-2}) and (\ref{ME-3}). We observe that the quasienergy (\ref{QE-3}) consists of a dynamical contribution plus a topological one determined by the parity of the integer number $\ell = (1/2\pi)\int_0^T \partial \varphi/\partial t \ {\rm d}t$, which accounts for the windings gathered by the FSSs around the north pole of the Bloch sphere. This motivates the introduction of an effective Berry phase $\phi_{\rm B}^{\rm eff} \equiv \ell \pi$ capturing the topological features of the FSSs.



\begin{thebibliography}{99}

\bibitem{SKGOKNSS13}
H. Sanada, Y. Kunihashi, H. Gotoh, K. Onomitsu, M. Kohda, J. Nitta, P. V. Santos, and T. Sogawa, Nature Phys. {\bf 9}, 280 (2013).  

\bibitem{NFSRN13} 
F. Nagasawa, D. Frustaglia, H. Saarikoski,  K. Richter, and J. Nitta, Nature Comm. {\bf 4}, 2526 (2013).

\bibitem{BMKNZ03}
A. Bohm, A. Mostafazadeh, H. Koizumi, Q. Niu, and J. Zwanziger, {\it The Geometric Phase in Quantum Systems} (Springer-Verlag, New York, 2003).

\bibitem{lyanda-geller} 
Y. Lyanda-Geller, Phys. Rev. Lett.  {\bf 71}, 657 (1993).

\bibitem{SVBNNF15}
H. Saarikoski, J.E. V\'azquez-Lozano, J.P. Baltan\'as, F. Nagasawa, J. Nitta, and D. Frustaglia, 
Phys. Rev. B {\bf 91}, 241406(R) (2015). 

\bibitem{SVBNF16} 
A classical equivalent was addressed in 
H. Saarikoski, J.P. Baltan\'as, J.E. V\'azquez-Lozano, J. Nitta, and D. Frustaglia, 
J. Phys.: Condens. Matter {\bf 28}, 166002 (2016).

\bibitem{G-LP12}
For a study on related topological aspects in driven systems, see
A. G\'omez-Le\'on and G. Platero, Phys. Rev B {\bf 86}, 115318 (2012). 

\bibitem{Kitagawa}
T. Kitagawa, E. Berg, M. Rudner and E. Demler, Phys. Rev. B {\bf 82}, 235114 (2010).

\bibitem{Lindner}
 N.H. Lindner, G. Refael, and V. Galitski, Nature Phys. {\bf 7}, 490 (2011).

\bibitem{CiracFloquet}
L. Jiang, T. Kitagawa, J. Alicea, A.R. Akhmerov, D. Pekker, G. Refael, J.I. Cirac, E. Demler, M.D. Lukin and P. Zoller, Phys. Rev. Lett. {\bf 106}, 220402 (2011).

\bibitem{FrustagliaFloquet}
A.A. Reynoso and D. Frustaglia, Phys. Rev. B {\bf 87}, 115420 (2013).

\bibitem{BS40}   
F. Bloch and A. Siegert, Phys. Rev. {\bf 57}, 522 (1940). 

\bibitem{berry} 
M. V. Berry, Proc. R. Soc. London A  {\bf 392}, 45 (1984).

\bibitem{PFR03}
M. Popp, D. Frustaglia, and K. Richter, Phys. Rev. B {\bf 68}, 041303(R) (2003).

\bibitem{aharonov} 
Y. Aharonov and J. Anandan, Phys. Rev. Lett.  {\bf 58}, 1593 (1987).

\bibitem{S65}
J. H. Shirley, Phys. Rev. {\bf 138}, B979 (1965).

\bibitem{F39}   
R. P. Feynman, Phys. Rev. {\bf 56}, 340 (1939).

\bibitem{S72}  
S. Stenholm, J. Phys. B: Atom. Molec. Phys. {\bf 5}, 890 (1972).

\bibitem{note-1} 
We choose the Floquet solution with $0<\varepsilon^+<\frac{\hbar\omega_0}{2}$ which is equivalent to the solutions with quasienergy $\varepsilon^+ + n \hbar\omega_0$. 

\bibitem{ABRPR} Y. Aharonov, E. Ben-Reuven, S. Popescu, and D. Rohrlich, Phys. Rev. Lett. {\bf 65}, 3065 (1990); {\it ibid.}, Nucl. Phys. B {\bf 350}, 818 (1991).

\bibitem{YGOC16}
Z.-J. Ying, P. Gentile, C. Ortix, and M. Cuoco, 
Phys. Rev. B {\bf 94}, 081406(R) (2016).

\bibitem{WS88}
W. S. Warren and M. S. Silver, in Advances in Magnetic and Optical Resonance, edited by J. S. Waugh (Academic Press, 1988), pp. 247-384.

\bibitem{SA16}
D. Suter and G. A. \'Alvarez, Rev. Mod. Phys. {\bf 88}, 041001 (2016).

\bibitem{OYLBLO05}
W. D. Oliver, Y. Yu, J. C. Lee, K. K. Berggren, L. S. Levitov, and T. P. Orlando, 
Science {\bf 310} 1653-1657 (2005).

\bibitem{note-2}
W. D. Oliver, private communication (2016).

\bibitem{LZS}
C. Barthel, D. J. Reilly, C. M. Marcus, M. P. Hanson, and A. C. Gossard, Phys. Rev. Lett. {\bf 103}, 160503 (2009); J. Stehlik, Y. Dovzhenko, J. R. Petta, J. R. Johansson, F. Nori, H. Lu, and A. C. Gossard, Phys. Rev. B {\bf 86}, 121303(R) (2012); G. D. Fuchs, G. Burkard, P. V. Klimov, and D. D. Awschalom, Nature Phys. {\bf 7}, 789 (2011); J. J. Pla, K. Y. Tan, J. P. Dehollain, W. H. Lim, J. J. L. Morton, F. A. Zwanenburg, D. N. Jamieson, A. S. Dzurak, and A. Morello, Nature {\bf 496}, 334 (2013); F. Forster, G. Petersen, S. Manus, P. H\"anggi, D. Schuh, W. Wegscheider, S. Kohler, and S. Ludwig, Phys. Rev. Lett. {\bf 112}, 116803 (2014).

\bibitem{BSRF17}
J.P. Baltan\'as, H. Saarikoski, A.A. Reynoso, and D. Frustaglia, arXiv:1703.07100.

\bibitem{FR01}
D. Frustaglia and K. Richter, 
Found. Phys. {\bf 31}, 399 (2001).

\end{thebibliography}
\end{document}